# Gaze-contingent decoding of human navigation intention on an autonomous wheelchair platform

Mahendran Subramanian[1,2], Suhyung Park[1], Pavel Orlov[1,2], Ali Shafti[1,2], A. Aldo Faisal[1,2]

*Abstract*— We have pioneered the Where-You-Look-Is Where-You-Go approach to controlling mobility platforms by decoding how the user looks at the environment to understand where they want to navigate their mobility device. However, many natural eye-movements are not relevant for action intention decoding, only some are, which places a challenge on decoding, the so-called Midas Touch Problem. Here, we present a new solution, consisting of 1. deep computer vision to understand what object a user is looking at in their field of view, with 2. an analysis of where on the object's bounding box the user is looking, to 3. use a simple machine learning classifier to determine whether the overt visual attention on the object is predictive of a navigation intention to that object. Our decoding system ultimately determines whether the user wants to drive to e.g., a door or just looks at it. Crucially, we find that when users look at an object and imagine they were moving towards it, the resulting eye-movements from this motor imagery (akin to neural interfaces) remain decodable. Once a driving intention and thus also the location is detected our system instructs our autonomous wheelchair platform, the A.Eye-Drive, to navigate to the desired object while avoiding static and moving obstacles. Thus, for navigation purposes, we have realised a cognitive-level human interface, as it requires the user only to cognitively interact with the desired goal, not to continuously steer their wheelchair to the target (low-level human interfacing).

## I. INTRODUCTION

Wheelchairs and exoskeletons provide independence to people suffering from the loss of mobility. For people with more severe disabilities, such as amyotrophic lateral sclerosis (ALS) or high-level spinal cord injuries interfaces such as joysticks are inadequate. Alternative interface methods include sip-and-puff, head-mount gyroscope-based devices, gaze-controlled graphical user interfaces (GUIs) and electroencephalography (EEG)-based steering. However, these methods involve a high cognitive load due to attention switching between interface and environment, and they often only allow low-level navigational input (e.g., requiring continuous steering commands to turn). Besides, clinicians have observed that nearly half of their patients who cannot use conventional wheelchair control methods would benefit from navigational assistance [1]. Hence, there is a need for better human interfacing in wheelchairs with autonomous driving capabilities.

In gaze-controlled wheelchair systems, the driver's eye movements are recorded and different processing methods of the gaze information result in different control options for the driver. Simple methods use the gaze direction and map it directly to wheelchair control commands, meaning that the eyes essentially act as a joystick [2,3]. These types of eye movements, however, may feel unnatural and strenuous for the driver, especially after a long period of use. Others have shown the use of gaze to select navigation options on a graphical user interface (GUI) [4,5]. In [6], fuzzy set theory is used to combine information obtained from the user's gaze fixations, and the surrounding environment, to initiate navigation. We have shown the use of supervised learning to infer directional intention from motor imagery-based gaze patterns [7]. Such interfaces which involve the user looking at their environment to initiate motion will be more comfortable for prolonged use, as opposed to cases requiring them e.g., to gaze upon a screen. An intuitive gaze interface would allow users to gaze upon their environment naturally and detect driving intention purely from natural gaze dynamics and the context of the environment. This brings us to the Midas Touch problem: not every part of the environment you gaze upon, is one you intend to act upon or drive to. It is therefore essential to build systems that can differentiate between natural gazing without an interactive intention and gaze that is the result of an intention of action [8]. The appeal of using natural behaviour predictive of future actions for assistive technology [7, 9-13] arises from the fact that users do not have to learn to operate a control system (e.g., by looking at a computer screen while being eye-tracked), or learn unnatural behaviours that facilitate decoding (e.g., staring at things to generate artificial long dwell times or perform gaze gestures), but simply behave normally.

Gaze patterns are known to vary based on the observer's intention [14], making it a source for information about the user's intentions. Classifiers based on machine learning methods have been shown to identify the task given to the observer (up to 70% accuracy for non-binary classification) based on saccade information: mainly spatial density, timing, and length of gaze [15]. In our case, the task of the wheelchair user is to select a target location via gaze. We propose to fulfil this task based on the user's natural gaze features by deploying machine learning methods to identify the context of the gaze, i.e., which object they are looking at and to analyse the visual attention upon that given object to predict the intention of navigation towards the object. Our gazeinformatics-based intention decoder is used to continuously decode the intentions of the wheelchair user, i.e., continuously detecting the navigational intentions they are trying to achieve, while our autonomous wheelchair platform, the A.Eye-Drive, is on the move and is accounting for any static and dynamic obstacles faced. This structure allows the user to merely provide destination commands via natural gaze, with the system decoding it, while the autonomous wheelchair keeps track of obstacles, and takes care of path planning and navigation. This significantly reduces the cognitive load for the user and

Brain and Behavior lab: [1]Dept. of Bioengineering, [2]Dept. of Computing, Imperial College London. We acknowledge funding from EPSRC [EP/N509486/1: 1979819], the Mobility Unlimited Challenge Discovery Award and a UKRI Turing AI Fellowship to AAF.

eliminates the need to interact with a wheelchair control interface all the time; replacing it with a natural interface instead.

## II. MATERIAL AND METHODS

### A. System architecture

Our system consists of two modules (I) The autonomous wheelchair platform consists of Real-time SLAM (Simultaneous Localisation and Mapping), navigation and wheelchair control. A two-dimensional online-SLAM algorithm is used to both create the map and perform localisation in real-time. The wheelchair localisation is enhanced by deploying Lidar-based odometry from the 2D SLAM. Scan match between consecutive laser scans is done to estimate the position of the lidar. Point clouds from both the 2D Lidar and the RGB-D camera are used for obstacle detection. Modified Dijkstra's algorithm and dynamic window approach are used for path planning and obstacle avoidance. (II) The autonomous navigation architecture was then incorporated with the gaze-based destination commands i.e., firstly upon classifying the user's gaze pattern, the natural gaze-based intention decoder module publishes predictor messages and object identifier messages, i.e., whether the user intends to interact with the object of interest within the field of view. Next, the gaze monitor subscribes to these messages, and the gaze-based commands are published. Finally, 3D gaze-based end-point control ('Wink' detection) was used as a double confirmation of the decoded high-level intention and an option for low-level intention (free navigation) input as well. Now the predicted intention has been converted to a goal based on the object of interest's location. Once a path to the goal has been computed, the required velocity commands are sent to the dual h-bridge, which is the driver for the wheel motors. And all these processes were achieved in real-time. Gaze-based intention decoding, and target localization are explained in the following subsections.

### B. Gaze informatics for intention decoding

We studied gaze fixations in different intention scenarios, to build an intention decoding engine. This module decodes the users' high-level intention. Subjects perform tasks with and without interactive intentions and the gaze tracking software tool records the corresponding gaze patterns. Classifiers are then trained on this data. To collect the data for model training and evaluation, we perform data collection sessions with 5 healthy subjects. During the sessions, each subject wears SMI eye-tracking glasses and performs 11-14 trials. We propose interactive observation tasks, in which participants are shown a target object and are asked to perform different actions with them (see Fig. 1). Orange crosshair is the relative 2D gaze (X, Y), binocular. Orange circle around the crosshair is the gaze fixation duration, i.e., how long the user is fixing his gaze at that position (X, Y). A computer provides voice commands based on these tasks for each subject and the gaze tracking software tool records the corresponding gaze patterns. Target objects used in the scope of this paper were {TV, Laptop and Chair}. Fig. 1. illustrates the bounding boxes that we used for object labelling. We then map the gaze point to object labels. From one image to another, the bounding box that is drawn for a single object might change because of head movement, different object locations, and different viewing points (see Fig. 1). To account for this, we average bounding boxes per object and normalize gaze point position within it. We collected approximately 3500-3700 gaze points for each object.

To predict a high-level intention, we used visual attention density to determine intention. To this end, we used a simple but robust approach: gaze locations within the object bounding box were normalised. This 2D location was fed to an object-specific Fine Gaussian Support Vector Machine (SVM) or Weighted K-Nearest Neighbours (KNN) that was trained on the two classes (10-fold cross-validated) i.e., Class 1 -- non-interactive tasks; Class 2 -- interactive tasks. Each frame classification output was fed into a ring-buffer of 40 frames, and we performed a winner-take-all vote to determine the so temporally averaged intention. The intention decoder pipeline can be seen in Fig. 2. The above process was done offline, but to achieve real-time intention decoding, Semantic Fovea was used to compute object label and bounding box [16].

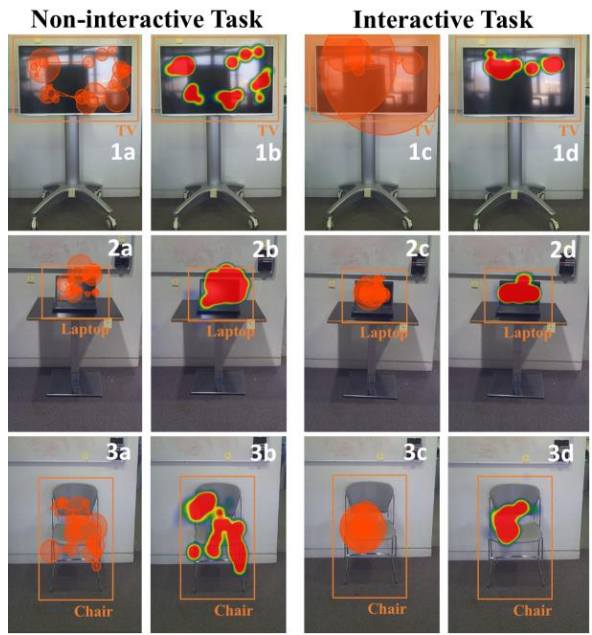

**Figure 1.** Gaze pattern comparison from labelled ego-centric images between non-interactive (left side) and interactive (right side) tasks for one subject and three objects: TV, laptop, and chair. The first and third columns show the saccade, where the orange crosshair and circle diameter represents gaze fixation and fixation duration, respectively. The second and fourth columns are heatmaps – Red (High intensity) to Green (Low Intensity). The tasks given were: (1a and 1b) - Look at the TV; (1c and 1d) - Look at the TV and imagine watching a movie; (2a and 2b) - Look at the laptop; (2c and 2d) - Look at the laptop and imagine reading a paragraph; (3a and 3b) - Look at the chair; (3c and 3d) - Look at the chair and imagine sitting on it.

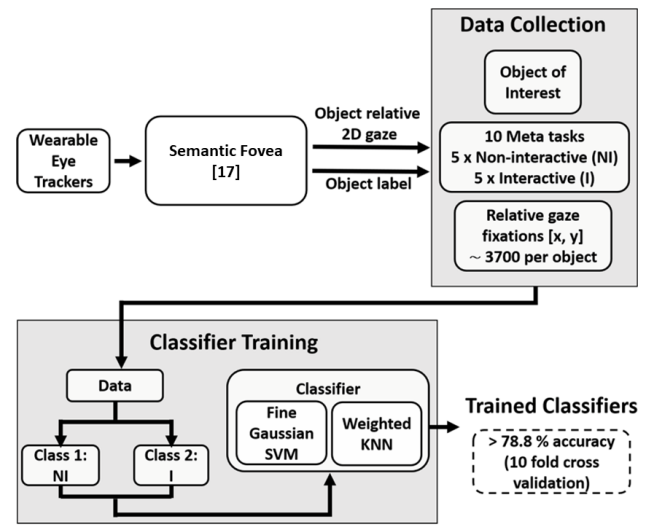

**Figure 2.** Natural gaze informatics-based intention decoder training pipeline.

### C. Integration

The intention decoder outputs the intention Boolean, and corresponding 3D gaze point coordinates, i.e., object of interest's location, is received from the 3D Gaze Estimation module (see Fig. 3). The 3D gaze Estimation procedure is explained in our earlier studies [7,17]. For the experimental setup in this paper, we communicate the detected intention to the user and ask them to confirm with a wink. This is done for safety reasons, as the system's performance is undergoing evaluation. Once the wink is detected on an interactive type intention, the 3D gaze point coordinates are sent to the navigation node as a goal. After a path to the intended destination has been planned based on the occupancy grid map, the required linear and angular velocity commands are sent to the wheelchair drive node - which is the driver for the wheel motors. These nodes are entirely native to this architecture. All of these processes were achieved online.

Using the integrated system, the user simply needs to look at the object within a room that they would like to interact with; the classifier will detect this intention, while the gaze monitor system stores the point in space that the wheelchair would have to go to. Once confirmed by the user through a wink, the wheelchair will autonomously navigate to the given point, avoiding obstacles and collisions. The above-described individual models and the integrated system are evaluated with results described in the next section.

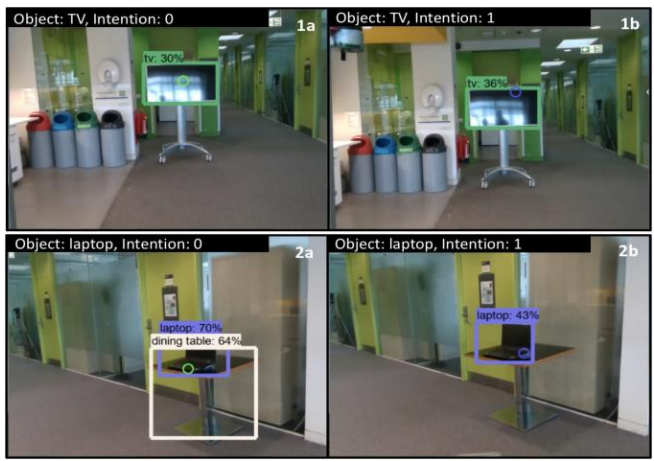

**Figure 3. Real-time gazeinformatics-based intention decoder.** Different coloured boxes denote the identified objects with object detection accuracy. The green circle denotes the gaze point (binocular) and non-interactive gaze class. It turns purple when an interactive type intention is decoded. Gaze on TV: Non-interactive (1a), Interactive (1b). Gaze on Laptop: Non-interactive (2a), Interactive (2b).

## III. RESULTS AND DISCUSSION

### A. Gazeinformatics-based intention decoder evaluation

The accuracy level results of the classifiers trained on gaze data is given below. For each object, we have listed the classifier that provides the best result and the result value. This is generally higher than 78.8%. In Fig. 1, we can observe a clear separation between the positions of gaze point for different tasks. As the classifiers are trained on natural human behavioural data, their usage can result in a more intuitive command interface. We did take different approaches when it came to the Machine Learning based Classifier. And chose the one which gave high classification accuracy after training. Often this accuracy for the classification techniques varied depending on the object and its features.

The classifier that gave the highest accuracy for the objects were: TV (Fine Gaussian Support Vector Machine – 78.80%), Laptop (Weighted K-Nearest Neighbor – 86.20%), and Chair (Weighted K-Nearest Neighbor – 84.80%). We chose SVM and KNN as they performed well with respect to others in most cases (objects) and were computationally inexpensive. We integrate the trained models within the real-time gaze monitoring system. After building the real-time intention decoder, we recruited 10 healthy subjects different from the ones whose gaze behaviour were used to train the classifier. We recorded the gaze movements for these new subjects for the same interactive and non-interactive tasks to be classified using our real-time intention decoder. Accuracy of the resulting intention Boolean (0: Non-interactive, 1: Interactive) was evaluated. Real-time intention decoder performance evaluation results with standard error provided in Fig. 4 is obtained from tests with 10 subjects (n=12). From the results, we see that classifiers generally perform well in detecting non-interactive intention but have varied performance when it comes to detecting interactive ones. From the trend in Fig. 4, we can infer that real-time interactive intention decoding accuracy is higher for objects with a higher number of visual features (i.e., see results for the chair as opposed to TV in Fig. 4). While earlier studies demonstrate gaze-based prediction of human intention (up to 80 % accuracy) as well as the desired object before the user could reach out for the said object [18], task prediction experimentation results does reveal that spatial position and local features are of importance while differentiating task categories [15].

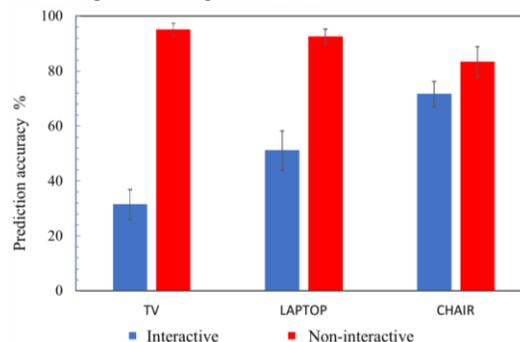

**Figure 4. Real-time gaze informatics-based intention decoder accuracy.**

### B. Integrated system evaluation

With the complete integrated system, we demonstrate that if, given a high-level cognitive intention (e.g., 'want to interact with the TV'), the wheelchair platform can successfully identify and act on low-level intentions of the driver, i.e., navigational intentions. The evaluation results are provided below. Every time users look at the objects: TV, Laptop and Chair (the list can be extended) with an interactive intention, the wheelchair autonomously navigates to the intended destination, i.e., to the object of interest's location. The results with standard error (n=6) from 2 subjects are provided as proof of concept. Experiments with further subjects are ongoing. During the evaluation of the intention decoding semi-autonomous wheelchair setup, around 29.1±1.97 seconds were taken to reach the goal during a single navigation run. The static obstacle detection and avoidance rate was 95±5% and the dynamic obstacle detection and avoidance rate was 90±6.12%. Most importantly, the average stop distance from the goal was about 24±9.79cm and the number of stops (in case of emergency) was lower than 1 i.e., 0.4±0.24. When compared with earlier studies [7, 17, 19] that tried to achieve natural gaze data-driven wheelchair navigation, the subjects seemed to perform faster with a low number of stops during navigation. This is understandable as the autonomous platform take over obstacle identification and low-level navigational intentions. Most of all the integrated system provides better control of the wheelchair with the introduction of object-specific interactive intention mediated goal definition as well as 3D gaze-based endpoint control. Which is much more intuitive than our previous gaze-based Natural Decoder [7], Continuous Control Field [7], and "Look where you Want To Go" [19] wheelchair interfaces.

IV. CONCLUSION

The technology presented here aims to tackle a fundamental challenge to independent mobility of severely disabled wheelchair users: the obligation to interact and provide continuous navigational commands. Eye movements are a natural human interface that are intuitive for users and tend to be retained in even the most severe types of motor impairments. Typical gaze interfaces for assistive devices tend to rely on unnatural gaze patterns, i.e., forcing the user to fixate for a period of time, or draw gaze gestures, to indicate a certain intention. Here, we present a solution in the form of a seamless and straightforward gaze interface, which allows the user to naturally gaze upon their desired destination and have their intention decoded, effectively overcoming the Midas Touch problem of identifying the relevant natural eye-movements for action intention decoding. The decoded intention is seamlessly communicated to our autonomous wheelchair platform, the A.Eye-Drive, which proceeds to take care of the tedious navigational work involving motion and path planning for static and dynamic obstacles.

Previous work has shown implementations of gaze-based intention decoding with more sophisticated formulations such as Hidden Markov Models and Partially Observable Markov Decision Processes [18, 20]. Here, we have demonstrated a robust, yet very computationally light classifier capable of class contingent real-time intention decoding from the wheelchair user's natural gaze. This system is a dynamic equivalent to the static histogram-based classifier that we proposed earlier [7] and presents an opportunity for further development by adding more objects and intention classes to the system allowing for more challenging scenarios during evaluation.